\documentclass{article}
\usepackage{spconf,amsmath,graphicx}
\usepackage{amssymb}
\usepackage{cite}
\usepackage{color}
\usepackage[table,xcdraw]{xcolor}
\usepackage{balance}

\let\OLDthebibliography\thebibliography
\renewcommand{\thebibliography}[1]{%
  \OLDthebibliography{#1}%
  \fontsize{9}{10}\selectfont
  \setlength{\parskip}{0pt}%
  \setlength{\itemsep}{0pt}%
  \setlength{\parsep}{0pt}%
  \setlength{\topsep}{0pt}%
}

\title{Beyond U-Net: A Latent-Representation-Aligned Skip-Free Backbone for Flow-Matching Speech Enhancement}

\name{Wangyi Pu$^\ast$ and Michele Scarpiniti\thanks{$^\ast$ Corresponding author: wangyi.pu@uniroma1.it.}}
\address{Department of Information Engineering, Electronics and Telecommunications,\\Sapienza University of Rome, Italy}

\begin{document}
\ninept
\maketitle
%

\begin{abstract}
Generative models, particularly diffusion and score-based approaches, have recently achieved strong performance in speech enhancement, but their iterative sampling process limits real-time deployment. Flow Matching offers an efficient alternative by transporting noisy speech toward clean speech through an ordinary differential equation with few function evaluations. In this work, we propose a skip-free encoder-decoder backbone for flow-matching speech enhancement, guided by Latent Representation Alignment (LRA). Instead of relying on U-Net skip connections, which may transfer noise-correlated low-level features to the decoder, the proposed model aligns its bottleneck and decoder representations with clean latent features extracted from a frozen Descript Audio Codec encoder-decoder without quantization. This codec-aligned supervision promotes compact clean-speech representations while preserving efficient few-step inference. Experiments on WSJ0-CHiME3 and VoiceBank-DEMAND show improved PESQ and perceptual quality, especially on VoiceBank-DEMAND, using only five function evaluations.
\end{abstract}
%

\section{INTRODUCTION}
\label{sec:intro}

The objective of speech enhancement (SE) is to recover clean speech signals from recordings corrupted by environmental noise \cite{loizou2007speech, hendriks2022dft}. In recent years, generative approaches, including GANs, normalizing flows, diffusion models, and score-based methods, have shown strong performance in SE \cite{baby2019sergan,nugraha2020flow,bie2022unsupervised,richter2023speech,lemercier2023analysing,lemercier2023storm,lay2023reducing,gonzalez2024diffusion,guo2024variance,lay2024single}. However, diffusion and score-based models typically require iterative reverse-time sampling, repeatedly evaluating a deep neural network (DNN) over many discretization steps. This often results in a high number of function evaluations (NFE), commonly larger than 25, which can hinder real-time deployment \cite{richter2023speech,lemercier2023analysing,lemercier2023storm,lay2023reducing,gonzalez2024diffusion,guo2024variance,lay2024single}. To reduce this inference cost, Flow Matching (FM) has recently been introduced for SE, as in FlowSE \cite{lipman2023flow,tong2024improving,liu2024generative,jung2024flowavse,lee2025flowse}. By learning a continuous normalizing flow governed by an ordinary differential equation (ODE) \cite{chen2018neural}, FM enables noisy speech to be transported toward clean speech through a simple probability path, achieving competitive enhancement quality with as few as five function evaluations.

Despite these inference speedups, training generative SE models can remain computationally demanding, with the backbone architecture playing a key role in convergence and final quality. Although U-Net skip connections preserve fine acoustic details, they may also transfer noise-correlated low-level features from the corrupted input to the decoder \cite{si2024freeu}. This can increase the burden on the decoder, which must simultaneously suppress residual noise and reconstruct clean speech, motivating the design of a skip-free backbone guided by clean latent representations.

To address this limitation, we propose a skip-free encoder-decoder backbone for FM-based SE. Rather than transferring encoder features through U-Net skip connections, the proposed model relies on Latent Representation Alignment (LRA), which supervises the bottleneck and decoder representations using clean latent features extracted from a frozen high-fidelity audio codec. In particular, we repurpose the Descript Audio Codec (DAC) by bypassing its residual vector quantization stage, thus using it as a continuous acoustic autoencoder for representation-level supervision during training.

By replacing structural shortcuts with prior-guided latent alignment, the proposed method mitigates the potential noise leakage of U-Net skip connections and promotes faster learning of perceptually relevant clean-speech representations. Experimental results show improved PESQ and competitive overall performance under an efficient inference setting with only five function evaluations.

\begin{figure}[t]
    \centering
    \includegraphics[width=\linewidth]{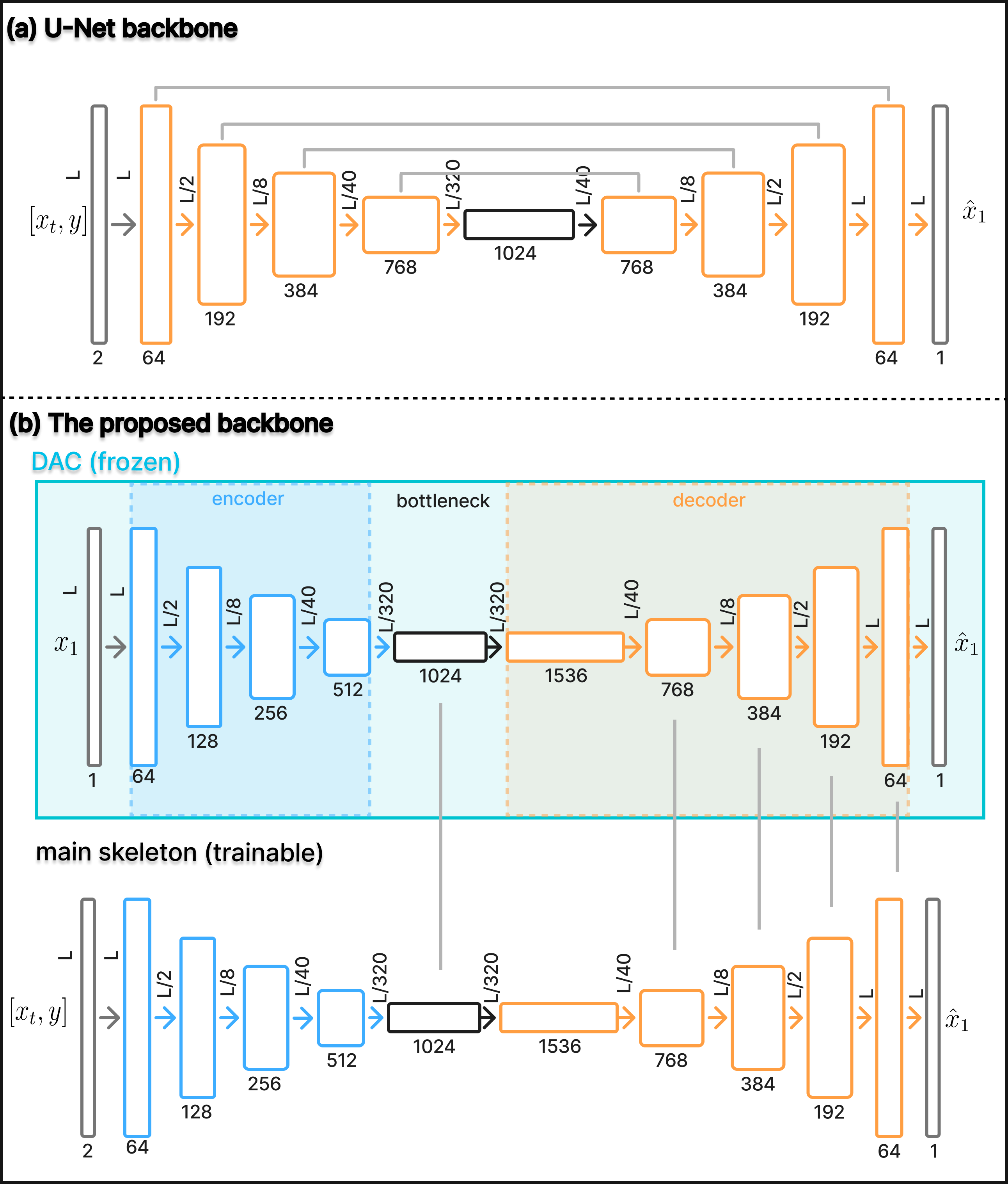}
    \caption{
        Overview of the proposed codec-aligned skip-free waveform backbone. 
        (a) The symmetric U-Net baseline takes $[x_t, y]$ as a 2-channel input and uses four skip connections. 
        (b) The proposed asymmetric skip-free backbone removes skip connections and predicts the complete clean waveform $\hat{x}_1$. A frozen DAC without RVQ provides clean latent targets for LRA.
    }
    \label{fig:backbone_overview}
\end{figure}

\section{BACKGROUND}
\label{sec:background}

\subsection{Problem Formulation}
\label{subsec:problem_formulation}

Speech enhancement aims to recover a clean speech signal $x_1 \in \mathbb{R}^L$ from a noisy speech $y = x_1 + n \in \mathbb{R}^L$, where $n$ represents additive environmental noise and $L$ denotes the temporal length. In the context of Flow Matching for SE, we define a continuous-time generative process over a time variable $t \in [0, 1]$. The process starts from a conditional prior centered around the noisy speech, $x_0 \sim p_0(x_0|y) = \mathcal{N}(y, \sigma^2\mathbf{I})$ \cite{lee2025flowse, richter2023speech}, where $\sigma \ge 0$, and evolves toward the clean speech sample $x_1$ at $t=1$.

\subsection{Flow Matching for Speech Enhancement}
\label{subsec:cfm}

The generative process is governed by an ordinary differential equation (ODE) defined by a time-dependent vector field $v_t(x_t|y)$:
\begin{equation}
\label{eq:ode}
\frac{d x_t}{dt}=v_t(x_t|y),
\qquad x_0\sim\mathcal{N}(y,\sigma^2\mathbf{I}).
\end{equation}
Following the FlowSE formulation, we define the conditional probability path $p_t(x_t|x_1,y)$ as a Gaussian distribution:
\begin{equation}
p_t(x_t|x_1,y) = \mathcal{N}\!\big(\mu_t(x_1,y),\sigma_t^2\mathbf{I}\big),
\end{equation}
whose mean linearly interpolates between the noisy and clean speech signals, while its standard deviation decreases linearly:
\begin{align}
\label{eq:mean_path}
\mu_t(x_1,y) &= t x_1+(1-t)y, \\
\label{eq:std_path}
\sigma_t &= (1-t)\sigma.
\end{align}
Accordingly, an intermediate state can be written as $x_t = \mu_t(x_1,y) + \sigma_t \epsilon$, where $\epsilon\sim\mathcal{N}(0,\mathbf{I})$. Conditional Flow Matching (CFM) trains a neural network $v_\theta(x_t,y,t)$ to match the target conditional vector field associated with this path:
\begin{equation}
\label{eq:cfm_loss}
    \mathcal{L}_{CFM}(\theta) = \mathbb{E}_{t, (x_1, y), p_0(x_0|y)} \! \left[ \left\| v_\theta(x_t, y, t) - v_t(x_t|x_1, y) \right\|_2^2 \right],
\end{equation}
where $t$ is sampled from $[0, 1]$ during the ideal formulation and from $[0, 1-t\delta]$ in practice to avoid numerical singularities near $t=1$.

For the Gaussian path above, the derivatives of the mean and standard deviation are:
\begin{equation}
\label{eq:derivatives}
\mu'_t(x_1,y)=x_1-y \quad \text{and} \quad \sigma'_t=-\sigma.
\end{equation}
The corresponding conditional vector field is:
\begin{equation}
\label{eq:general_vf}
v_t(x_t|x_1,y) = \mu'_t(x_1,y) + \frac{\sigma'_t}{\sigma_t} \big(x_t-\mu_t(x_1,y)\big).
\end{equation}
Substituting derivatives in \eqref{eq:derivatives} into \eqref{eq:general_vf} yields the closed-form target field:
\begin{equation}
\label{eq:target_vector_field}
v_t(x_t|x_1,y) = x_1-y-\sigma\epsilon = \frac{x_1-x_t}{1-t}, \quad t<1.
\end{equation}
At inference time, the enhanced speech signal is obtained by numerically solving the ODE from $t=0$ to $t=1$. Starting from $x_{t_0}\sim\mathcal{N}(y,\sigma^2\mathbf{I})$, the interval $[0,1]$ is discretized into $N$ steps ($0=t_0<\cdots<t_N=1$), and the Euler update is:
\begin{equation}
\label{eq:euler_update}
x_{t_i} = x_{t_{i-1}} + v_\theta(x_{t_{i-1}},y,t_{i-1})\Delta t_i, \quad i=1,\ldots,N,
\end{equation}
where $\Delta t_i = t_i-t_{i-1}$. The final estimate is given by $\hat{x}_1 = x_{t_N}$.

\color{black}
\section{PROPOSED METHOD}
\label{sec:method}


U-Net skip connections preserve fine acoustic details by transferring multi-scale encoder features to the decoder but may also carry noise-correlated low-level features from the corrupted input. Motivated by this observation, we propose a skip-free encoder-decoder architecture for flow-matching-based SE. Instead of using structural skip connections, the proposed model is guided by LRA, a training-time representation-level supervision mechanism based on a frozen DAC prior. Unlike STFT-domain generative SE methods \cite{richter2023speech,lee2025flowse}, our framework operates directly on time-domain waveforms.

\subsection{$x$-Prediction Formulation}
\label{subsec:formulation}

The target vector field in \eqref{eq:target_vector_field} motivates an $x$-prediction parameterization, where the network directly estimates the clean waveform at each timestep $\hat{x}_1=f_\theta(x_t,y,t)$. The corresponding velocity used for ODE integration is then recovered as:
\begin{equation}
\label{eq:v_reparam}
v_\theta(x_t,y,t)=\frac{\hat{x}_1-x_t}{1-t}.
\end{equation}
This parameterization preserves the flow-matching sampling procedure while enabling direct waveform-domain, adversarial, and representation-level supervision. In practice, we train the model using an unweighted clean-speech prediction loss, which provides a stable alternative to directly regressing the velocity field near $t=1$.

\subsection{Skip-free Backbone}
\label{subsec:codec_aligned_backbone}

The proposed backbone consists of a trainable encoder and decoder without lateral skip connections (see Fig. \ref{fig:backbone_overview}). To facilitate latent representation alignment, the encoder follows the topology of the DAC encoder \cite{kumar2023high}, with the first layer modified to accept the two-channel input $[x_t,y]\in\mathbb{R}^{2\times L}$. The signal is processed through a channel progression from $2$ to $1024$ using DAC-style residual units, Snake activations, and strided convolutions with downsampling factors $[2,4,5,8]$. Feature-wise Linear Modulation (FiLM) layers \cite{perez2018film} are used to inject timestep embeddings into the network. The resulting bottleneck representation is denoted as $h_{\mathrm{bn}} \in \mathbb{R}^{1024\times L/320}$.

The decoder follows the corresponding DAC decoder topology and is initialized from the pre-trained DAC decoder. During training, this decoder is updated as part of the enhancement network. The residual vector quantization (RVQ) stage of DAC is bypassed, so that the architecture operates as a continuous acoustic autoencoder rather than a discrete codec. This design provides a natural correspondence between the trainable enhancement network and the frozen DAC encoder-decoder used as a representation teacher.

\subsection{Latent Representation Alignment}
\label{subsec:lra}

Removing skip connections increases the information burden on the bottleneck representation. LRA addresses this issue by aligning the internal representations of the enhancement network with clean-speech representations extracted from a frozen DAC model. Let $\Phi_{\mathrm{enc}}$ and $\Phi_{\mathrm{dec}}$ denote the frozen DAC encoder and decoder, respectively. Given the clean waveform $x_1$, we first compute the clean codec latent representation $z_1=\Phi_{\mathrm{enc}}(x_1)$. The bottleneck alignment loss is then defined as:
\begin{equation}
\mathcal{L}_{\mathrm{bn}}=
\mathbb{E}_{t,x_1,y}\!
\left[
\left\|P_{\mathrm{bn}}(h_{\mathrm{bn}},t)-z_1\right\|_2^2
\right],
\end{equation}
where $P_{\mathrm{bn}}$ is a time-aware projection conditioned on the timestep embedding via FiLM.

We also distill decoder features from the frozen DAC decoder:
\begin{equation}
\mathcal{L}_{\mathrm{dec}}=
\frac{1}{K}\sum_{k=1}^{K}
\mathbb{E}_{t,x_1,y}\!
\left[
\left\|P_{\mathrm{dec}}^{(k)}(d_\theta^{(k)})
-\Phi_{\mathrm{dec}}^{(k)}(z_1)\right\|_2^2
\right],
\end{equation}
where $P_{\mathrm{dec}}^{(k)}$ denotes a learnable projection head applied to the $k$-th decoder feature prior to alignment, implemented as a pointwise one-dimensional convolution.

The overall LRA objective combines these terms: 
\begin{equation}
\mathcal{L}_{\mathrm{LRA}}=
\mathcal{L}_{\mathrm{bn}}+\eta\mathcal{L}_{\mathrm{dec}},
\end{equation}
where $\eta$ controls the decoder alignment strength.

\subsection{Training Objective}
\label{subsec:objective}

The main prediction loss is the unweighted clean-waveform reconstruction loss:
\begin{equation}
    \mathcal{L}_{x}=
    \mathbb{E}_{t,x_1,y}\!
    \left[
    \left\|f_\theta(x_t,y,t)-x_1\right\|_2^2
    \right].
\end{equation}
We further use DAC-style multi-period and multi-resolution discriminators with adversarial loss $\mathcal{L}_{\mathrm{adv}}$ and feature matching loss $\mathcal{L}_{\mathrm{feat}}$ \cite{kumar2023high}. The full objective is:
\begin{equation}
    \mathcal{L}_{\mathrm{total}}=
    \lambda_x\mathcal{L}_{x}
    +\lambda_{\mathrm{lra}}\mathcal{L}_{\mathrm{LRA}}
    +\lambda_{\mathrm{adv}}\mathcal{L}_{\mathrm{adv}}
    +\lambda_{\mathrm{feat}}\mathcal{L}_{\mathrm{feat}}.
\end{equation}

\begin{table*}[t]
\centering
\caption{Speech Enhancement performance of FlowSE variants on the WSJ0-CHiME3 and VB-DMD datasets. DAC is reported as a clean-audio reconstruction reference only and is not included in best-score highlighting.}
\label{tab:performance}
\renewcommand{\arraystretch}{1.15}
\resizebox{\textwidth}{!}{
\begin{tabular}{lccccccccccc}
\hline
\multicolumn{12}{c}{Trained and Tested on WSJ0-CHiME3} \\
\hline
Method & NFE & PESQ & DNSMOS & WVMOS & ESTOI & SIG & BAK & OVRL & SI-SDR & SI-SIR & SI-SAR \\
\hline
DAC                         & -- & $2.36 \pm 0.25$ & $3.84 \pm 0.19$ & $3.53 \pm 0.30$ & $0.83 \pm 0.02$ & $3.35 \pm 0.18$ & $4.01 \pm 0.16$ & $3.03 \pm 0.21$ & $-22.95 \pm 8.29$ & -- & -- \\
\hline
FlowSE (U-Net w/o GAN loss) & 5 & $2.94 \pm 0.54$ & $3.84 \pm 0.24$ & $3.87 \pm 0.42$ & $\mathbf{0.93 \pm 0.05}$ & $3.56 \pm 0.08$ & $4.17 \pm 0.04$ & $3.33 \pm 0.10$ & $\mathbf{19.48 \pm 4.23}$ & $31.30 \pm 4.68$ & $\mathbf{19.84 \pm 4.35}$ \\
FlowSE (U-Net)              & 5 & $3.05 \pm 0.50$ & $3.98 \pm 0.20$ & $\mathbf{4.08 \pm 0.34}$ & $\mathbf{0.93 \pm 0.05}$ & $\mathbf{3.62 \pm 0.08}$ & $\mathbf{4.19 \pm 0.04}$ & $\mathbf{3.38 \pm 0.10}$ & $19.10 \pm 4.09$ & $30.22 \pm 4.48$ & $19.52 \pm 4.21$ \\
FlowSE (LRA, proposed)      & 5 & $\mathbf{3.06 \pm 0.46}$ & $\mathbf{4.01 \pm 0.18}$ & $3.87 \pm 0.30$ & $\mathbf{0.93 \pm 0.05}$ & $3.61 \pm 0.08$ & $4.17 \pm 0.05$ & $3.37 \pm 0.10$ & $17.31 \pm 2.94$ & $\mathbf{32.62 \pm 5.20}$ & $17.47 \pm 2.94$ \\
\hline
\multicolumn{12}{c}{Trained and Tested on VB-DMD} \\
\hline
DAC                         & -- & $2.97 \pm 0.29$ & $3.57 \pm 0.26$ & $4.22 \pm 0.27$ & $0.85 \pm 0.03$ & $3.47 \pm 0.18$ & $4.02 \pm 0.14$ & $3.17 \pm 0.21$ & $-13.19 \pm 7.52$ & -- & -- \\
\hline
FlowSE (U-Net w/o GAN loss) & 5 & $2.72 \pm 0.60$ & $3.38 \pm 0.31$ & $4.23 \pm 0.36$ & $0.86 \pm 0.10$ & $3.45 \pm 0.17$ & $3.98 \pm 0.19$ & $3.15 \pm 0.22$ & $\mathbf{18.75 \pm 3.65}$ & $\mathbf{31.08 \pm 7.54}$ & $\mathbf{19.36 \pm 3.52}$ \\
FlowSE (U-Net)              & 5 & $2.88 \pm 0.59$ & $3.46 \pm 0.29$ & $4.37 \pm 0.30$ & $\mathbf{0.87 \pm 0.09}$ & $3.49 \pm 0.17$ & $3.97 \pm 0.20$ & $3.17 \pm 0.21$ & $18.35 \pm 3.49$ & $28.42 \pm 6.14$ & $19.19 \pm 3.49$ \\
FlowSE (LRA, proposed)      & 5 & $\mathbf{3.11 \pm 0.63}$ & $\mathbf{3.51 \pm 0.29}$ & $\mathbf{4.41 \pm 0.28}$ & $\mathbf{0.87 \pm 0.09}$ & $\mathbf{3.50 \pm 0.15}$ & $\mathbf{4.00 \pm 0.17}$ & $\mathbf{3.19 \pm 0.20}$ & $16.87 \pm 2.47$ & $28.07 \pm 5.53$ & $17.49 \pm 2.41$ \\
\hline
\end{tabular}
}
\end{table*}

\section{EXPERIMENTS}
\label{sec:experiments}

\subsection{Datasets}

We evaluate our model on two datasets. The WSJ0-CHiME3 dataset mixes clean Wall Street Journal (WSJ0) speech \cite{wsj0} with CHiME3 \cite{chime3} noises at SNRs drawn from a uniform distribution between 0 and 20~dB, resulting in 12,777 training, 1,206 validation, and 615 test files.\footnote{https://github.com/sp-uhh/sgmse} The publicly available VoiceBank-DEMAND (VB-DMD) dataset mixes VCTK speech \cite{VCTK} with DEMAND noises. Training SNRs are $\{0, 5, 10, 15\}$ dB, and test SNRs are $\{2.5, 7.5, 12.5, 17.5\}$ dB. The training dataset is split into a training and validation dataset, while speakers ``p226'' and ``p287'' are reserved for validation \cite{richter2023speech,lemercier2023storm,lay2024single}.

\subsection{Experimental Setup}
\label{sec:setup}

\textbf{Model Variants \& Ablations:} We evaluate three variants of our FlowSE framework: FlowSE (U-Net w/o GANLoss), FlowSE (U-Net), and FlowSE (LRA). 
The U-Net variants use a standard U-Net backbone, whereas FlowSE (LRA) adopts our codec-aligned skip-free backbone, as shown in Fig~\ref{fig:backbone_overview}. 
All variants directly predict the clean speech $\hat{x}_1 = f_\theta(x_t, y, t)$. 
A continuous DAC variant without codebook quantization is reported separately as a clean-audio reconstruction reference, rather than as an SE baseline.

\textbf{Hyperparameters:} Models are trained on 2-second segments (32,000 samples) with a batch size of 8. 
We use Adam optimization \cite{kingma2014adam} with a learning rate set to $10^{-4}$. 
For adversarially trained variants, the same learning rate is used for both the generator and discriminator. 
The Exponential Moving Average (EMA) decay is 0.999. 
The flow-matching prior noise standard deviation is $\sigma = 0.487$, and the integration margin is set to $t_\delta = 0.03$ to prevent numerical instability. All models are trained until the validation PESQ and SI-SDR curves plateau.

\textbf{Loss Weights:} Corresponding to Section \ref{subsec:objective}, the waveform prediction weight is set to $\lambda_x = 0.1$ for all FlowSE variants. 
For FlowSE (U-Net w/o GANLoss), adversarial and feature-matching losses are disabled, i.e., $\lambda_{\mathrm{adv}} = \lambda_{\mathrm{feat}} = 0$. 
For FlowSE (U-Net) and FlowSE (LRA), we set $\lambda_{\mathrm{adv}} = 1.0$ and $\lambda_{\mathrm{feat}} = 2.0$. 
For FlowSE (LRA), the overall LRA weight is $\lambda_{\mathrm{lra}} = 1.0$ and the decoder-level strength is $\eta = 1.0$.

\subsection{Evaluation Metrics}

Performance is assessed using WB-PESQ \cite{pesq}, Deep Noise Suppression MOS (DNSMOS) \cite{dnsmos}, wav2vec MOS (WVMOS) \cite{wvmos}, Extended Short-Time Objective Intelligibility (ESTOI) \cite{estoi}, DNSMOS P.835 (SIG, BAK, OVRL) \cite{ovrl}, and Scale-Invariant metrics (SI-SDR, SI-SIR, SI-SAR) \cite{sdr}.

\begin{figure}[htbp]
    \centering
    \includegraphics[width=\linewidth]{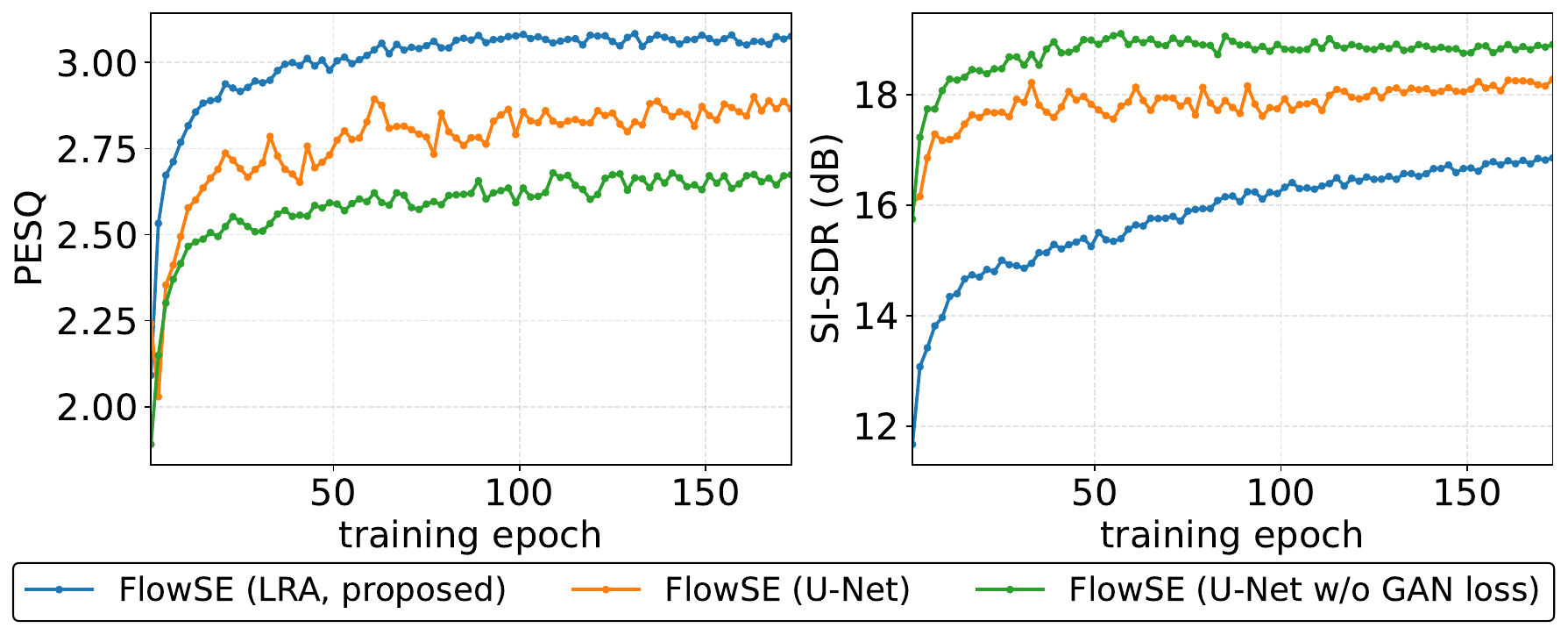} 
    \caption{Evolution of PESQ (left) and SI-SDR (right) metrics over training epochs on the test dataset of VB-DMD}
    \label{fig:training_curves}
\end{figure}

\section{NUMERICAL RESULTS}
\label{sec:results}

Table~\ref{tab:performance} and Fig.~\ref{fig:training_curves} summarize the quantitative results and training dynamics of the considered FlowSE variants.

\textbf{Effect of adversarial training:} Comparing FlowSE (U-Net w/o GAN loss) with FlowSE (U-Net), adversarial training improves perceptual quality on both datasets. On WSJ0-CHiME3, PESQ increases from $2.94$ to $3.05$, while WVMOS improves from $3.87$ to $4.08$. A similar trend is observed on VB-DMD, where PESQ increases from $2.72$ to $2.88$ and WVMOS from $4.23$ to $4.37$. This improvement in perceptual metrics is accompanied by a slight reduction in SI-SDR-related scores. Such a trade-off is consistent with the behavior of adversarial objectives, which tend to favor perceptual realism over strict waveform-level reconstruction.

\textbf{Effect of LRA:} Replacing the U-Net backbone with the proposed skip-free LRA backbone leads to clear perceptual improvements on VB-DMD and competitive performance on WSJ0-CHiME3. On VB-DMD, FlowSE (LRA, proposed) improves PESQ from $2.88$ to $3.11$ with respect to FlowSE (U-Net), while also achieving the best DNSMOS, WVMOS, SIG, BAK, and OVRL scores among the FlowSE variants. On WSJ0-CHiME3, the gain is more moderate: LRA slightly improves PESQ and DNSMOS, while the U-Net baseline remains better in WVMOS and SI-SDR-related metrics. These results suggest that LRA is particularly beneficial when the codec-derived clean latent prior provides an effective representation of the target speech distribution.

\textbf{Training dynamics:} Fig.~\ref{fig:training_curves} shows that the proposed LRA backbone reaches high perceptual quality with fewer training epochs than the U-Net baseline. This suggests that aligning the bottleneck and decoder features with clean codec representations provides useful representation-level guidance during training. In contrast to structural skip connections, which may also transfer noise-correlated encoder features, LRA encourages the model to rely on compact clean-speech representations. 

\textbf{Role of the DAC reference:} The DAC row is reported only as a clean-audio reconstruction reference and should not be interpreted as a speech-enhancement baseline. Its performance indicates the quality of the frozen codec prior used for representation alignment. On VB-DMD, where the DAC reconstruction metrics are stronger, LRA brings a larger improvement over the U-Net backbone. On WSJ0-CHiME3, the weaker DAC reconstruction metrics are consistent with the smaller gains observed for LRA. Importantly, FlowSE (LRA, proposed) outperforms the DAC reference in PESQ on both datasets, indicating that the proposed model does not merely reproduce the codec output, but combines codec-guided representation learning with the flow-matching enhancement objective.

\color{black}
\section{CONCLUSION}
\label{sec:conclusion}

In this paper, we proposed a codec-aligned skip-free encoder-decoder backbone for flow-matching-based speech enhancement. The proposed architecture removes conventional U-Net skip connections and compensates for the resulting information bottleneck through Latent Representation Alignment with a frozen continuous audio codec prior. By aligning bottleneck and decoder features with clean codec representations, the model is encouraged to learn compact clean-speech representations rather than relying on low-level structural shortcuts. Experiments on WSJ0-CHiME3 and VoiceBank-DEMAND show that the proposed method achieves competitive or improved perceptual quality under an efficient inference regime with only five function evaluations. 
Future work will investigate adaptive weighting of the alignment losses, extension to complex STFT-domain enhancement, and more detailed analysis of the relationship between codec latent quality and enhancement performance.

\clearpage
\balance
\bibliographystyle{IEEEbib}
\bibliography{refs}

\end{document}